# Exploration of User Privacy in 802.11 Probe Requests with MAC Address Randomization Using Temporal Pattern Analysis

Tomas BRAVENEC, Joaquín TORRES-SOSPEDRA, Michael GOULD, Tomas FRYZA

Institute of New Imaging Technologies • Universitat Jaume I • Avenida Sos Baynat, s/n • 12071, Castellón de la Plana, Spain

Department of Radio Electronics • Brno University of Technology • Technicka 12 • 61600, Brno, Czech Republic

Centro Algoritmi (CALG) • University of Minho • Av. da Universidade • 4800-058, Guimarães, Portugal

Tel.: (+34) 964 72 92 35 • E-Mail: bravenec@uji.es

**Summary:** *Wireless networks have become an integral part of our daily lives and lately there is increased concern about privacy and protecting the identity of individual users. In this paper we address the evolution of privacy measures in Wi-Fi probe request frames. We focus on the lack of privacy measures before the implementation of MAC Address Randomization, and on the way anti-tracking measures evolved throughout the last decade. We do not try to reverse MAC address randomization to get the real address of the device, but instead analyse the possibility of further tracking/localization without needing the real MAC address of the specific users. To gain better analysis results, we introduce temporal pattern matching approach to identification of devices using randomized MAC addresses.*

## Introduction

The technology that changed our lives the most in last decade is without question the introduction of the smartphone. Having internet connection on our person as we move through the world has had major impact on both our professional and personal lives. The practically constant connection to the internet, be it through cellular data or Wi-Fi, introduces a question of how much privacy, locational and otherwise, are people giving up. They are often giving up their privacy willingly for the use of services that make their daily lives easier; in other cases they do not know who or what may be identifying and/or tracking them.

Devices using Wi-Fi (also called WLAN) for connecting to the internet are extremely common, with most people having at least one around them at all times, for example mobile phone, smartwatch, laptop, and smart TVs. Since the majority of these devices are connected to the internet through wireless networks, the issue of privacy and device tracking on those networks should be something people are aware of. Our devices are communicating with the surrounding world using standardized protocols. For instance, a device in a IEEE 802 network is uniquely identified by the Media Access Control (MAC) address, which is used in all the messages involving the device. The device probe request is a type of wireless frame used to gather information about Wi-Fi access points in the proximity of a device. This is beneficial to the users as the device can identify and connect to a known access point without any user input, to switch to anohter AP with better coverage in a large public Wi-Fi network, as well as help with increasing accuracy of geolocation navigation by checking nearby Wi-Fi devices and comparing the signal strenght of detected access points with previously detected ones at the same location. These probe requests can be a major weak point of a Wi-Fi protocol, since they allow for non-cooperative user tracking if the device does not use enough privacy measures such as MAC address randomization.

Tracking using Wi-Fi protocols can vary as they can be used to determine the past whereabouts of users, current presence, or both. The past locations of devices can be determined if the devices are transmitting the preferred network list (list of the networks the device was connected to in the past) which can be matched to location using access point databases like WIGLE (2022). The current presence tracking can be done using a fingerprinting approach or in the case of devices without randomized MAC addresses, just by matching the globally unique MAC addresses of separate probe requests.

At the time of writing, most of the major operating systems have implemented some kind of privacy protection measures, that are helping to protect users from non-cooperative tracking. But the implementations and efficiency of those measures vary.

In this paper we explore the current state of privacy related measures in probe requests. We analyse how the situation has changed since the period before MAC randomization was first introduced, and we propose additional measures to further increase privacy. The main contribution of this paper is in the new angle of analysing the probe request datasets from the temporal point of view. We present a temporal pattern matching approach to identifying devices with randomized MAC addresses through the pattern of their appearances in time.

**Related Work**

The tracking of mobile device users using passive sniffing of probe requests has been a focus of research for quite a while now. For example Musa & Eriksson (2012) used probe requests for urban mobility tracking. The privacy vulnerability of the probe request frames was already proven in several publications by Ningning et al. (2013) or by Cunche et al. (2014) prior to the implementation of MAC address randomization. After the introduction of MAC address randomization in iOS 8 in 2014 (Vasilevsky et al., 2019), researchers worked to determine the inner working of the randomization technique Apple used (Freudiger, 2015). In other research the authors focused on determining the real MAC addresses assigned by each manufacturer (Martin et al., 2016). Di Luzio et al. (2016) determined the origin of people at large events using probe requests collected at 2 political events before elections in Italy and results were matching the official voting reports. Matte et al. (2016) provided details on bypassing MAC randomization with the use of temporal analysis, by exploiting the device specific timings between subsequent probe requests or scan instances. Martin et al. (2017) created very deep study of MAC address randomization and explored all the times it fails. Gu et. al. (2020) proposed an encryption for 802.11ac devices.

**Current Implementation of MAC Randomization**

Although the IEEE Standards Association Standards Board specified in 2018 a standard amendment 802.11aq-2018 (IEEE, 2018) considering randomization of MAC addresses, there is still no standard for actual implementation of randomization. This means each and every manufacturer and software developer can decide how to implement it in their own manner.

Addresses can be assigned either by the manufacturer or locally by the device network controller. The way the address is assigned is differentiated by the 2nd least significant bit of the first byte of the MAC address B1 as shown in Figure 1. If the bit is set, the MAC address was assigned locally. The least significant bit of the first byte B0 describes if the MAC address is unicast or multicast. For the majority of devices, this bit is set to 0. The unicast/multicast bit is set to 0 for individual devices and only set to 1 for device groups. This makes the distinction between globally unique and randomized MAC address of individual devices very simple as the 2nd digit of randomized MAC address in hexadecimal format can only be 2 (0010), 6 (0110), A (1010) or E (1110).

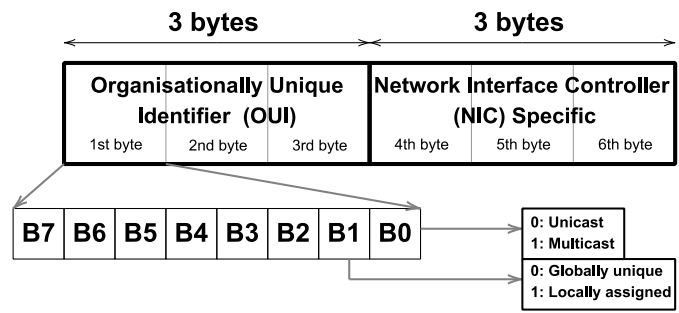

**Fig. 1:** Structure of MAC address with the functional bits

**Dataset**

We decided to collect a dataset to perform the analsysis. As we work with real data containing personal information, it stays to say that we only approached the analysis from the implementation of Wi-Fi protocol, specifically to explore potential privacy issues in current implementations. The data collection was limited to only passive capture and store of 802.11 management frames which we then anonymized before starting with analysis. The annonymization is done by hashing last 3 bytes of the MAC address, preffered network list, UUID-E and several other fields containing user specific information. From the anonymized data we do not observe personally identifiable information. Even though the collected data contains both real and randomized MAC addresses, it is not possible for us to match a MAC addresses to specific individuals because the analysis was done over annonymized version of the dataset, and we did not collect the data with presence in the office or the building.

To base our research work on up to date data, we collected probe requests at our small office for 6 days in December 2021. The office is in the corner of the $5^{th}$ floor and during peak times is occupied by about 15 researchers.. In that time we collected 340,360 probe requests. The data was collected using an ESP32 micro controller with connected micro SD card for storage of the collected probe requests. The firmware created for the ESP32 micro controller to capture probe requests and save them in standardized way readable by the Wireshark and similar packet analysis tools is publicly available from a GitLab repository (Bravenec, 2021).

About 10% of captured probe requests contained WPS (Wi-Fi Protected Setup) sections, which provide additional information about the device, starting with device name, manufacturer, and model. Since many devices use the name of their owner, this may pose a privacy leak in devices transmitting this additional information which is unnecessary for correct functioning of probe request frames. Even sending a device manufacturer name can reveal the user identity if the device itself is less common than others (e.g. Susie is the only Motorola user here). The most important issues of this WPS section, though, is the presence of UUID-E (Universally Unique IDentifier-Enrollee) data which is unique for a device since it is acquired using the globally unique MAC address of the device and does not change. Devices transmitting probe requests containing UUID-E are then easily localized as their globally unique MAC address can be recovered using UUID-E reversal techniques - by looking up the globally unique MAC address from hash tables (Martin et al., 2016). Therefore, we have hashed that additional information provided in the WPS Section to ensure privacy of users.

**Analysis**

In the past, the tracking of mobile devices using only probe requests was not very difficult as there were several factors that made identification of a single device fairly straightforward. These include nonrandomized MAC addresses, consecutive Sequence Numbers, common time difference between 2 probe requests, or extended information in the Information Element like supported transfer rates and vendor information.

*MAC addresses*

Even though MAC addresses cannot be used effectively to locate most modern devices, they can still be used to identify a single device during a single scan. From the analysis of the data collected at our office, the devices do not randomize MAC addresses after every probe request. This makes identification of a single scan instance from one device very easy, since they keep the same address for the scanning sequence, or multiple sequences. A solution to increase privacy would be to randomize the MAC address for every probe request, or at least more often than the devices do at the moment.

*Sequence Numbers*

Sequence numbers in probe request packets allow for another opportunity to easily identify packets coming from a single device during one scan instance, without the need to check the MAC address. The reason for this is the incremental nature of sequence numbers in probe requests coming from a single device. Every time the device sends a packet, the sequence number increments by 1. The sequence number can increase by more than 1, which happens if a device sends another packet or frame between 2 subsequent probe requests.

Addressing this issue would be fairly straightforward by using random sequence number for each probe request. This, combined with randomization after every probe request, would make identification of packets coming from a single device a lot more challenging as new techniques for identification through probes would be required.

*Fingerprinting with Information Elements*

To identify devices we collect the device specific information fields available in probe requests, out of which we create a single unique identifier (Loh et al. 2008). The information element fields in probe requests can contain various additional data, starting with supported transfer speeds, information about the vendor of the wireless chip inside of the device, and including the connected peripherals and device name. As mentioned in the section describing the dataset, WPS information might also be present, which contains enough information to create a unique fingerprint of the device. The biggest issue there is the UUID-E field which is unique per device and makes MAC address randomization pointless in devices that transmit WPS data, since instead of MAC address the UUID-E can be used to identify a device. And that is without considering UUID-E reversal techniques which can be used to determine the globally unique MAC address of the device (Martin et al., 2016).

For fingerprint creation we use all of the fields that remain constant for one device between transmissions. Supported transmission speeds, vendor information, WPS field and others are used to create a hash using the SHA512 algorithm. This ensures we have an unique fingerprint for information element of all devices. All of the fields in our device fingerprint are presented in Table 1, with the frequency of occurrences in the data collected at our office. As can be seen, the supported data rates are presented in 100% of collected probe requests, with extended list of supported data rates being missing from just 0.05% of the probes. The HT Capabilities (802.11n specific information regarding supported frequency bandwidth etc.) were also present in a majority of probe requests, followed up by extended capabilities and at least 1 vendor specific field, though the most common number of vendor specific fields for one probe request in the data collected in our lab was 4, in about 30 % of all probe requests. Since devices from the same vendor will have the same vendor specific fields, having 4 vendor specific fields the same, increases the probability that the devices are the same.

*Fingerprinting SSID lists*

By using previously mentioned techniques, it is easy to differentiate all probe requests sent by a single device in a single scan instance. Knowing that all probe requests came from a single device then allowed us to list all the different SSIDs in those probe requests. By using sets with each SSID represented only once, we can use set similarity as in equation (1) to calculate a probability of two devices being in fact a single device by using the transmitted SSID list.

$$p = \frac{set(A) \text{ and } set(B)}{set(A) \text{ or } set(B)} \qquad (1)$$

There is also a possibility for the attacker to identify the users directly through the SSIDs from the preffered network list, as there is a possibility to match some of the networks directly to people (for example SSID of network at university in another country, while we know our collegue is the only one around that used to study there).

| Information Element | Included in Probes | [%] |
|---|---|---|
| Supported Rates | 340360 | 100.00 |
| Extended Supported Rates | 340198 | 99.95 |
| HT Capabilities | 312227 | 91.73 |
| VHT Capabilities | 20252 | 5.95 |
| Extended Capabilities | 232918 | 68.43 |
| Vendor Specific Elements | 194801 | 57.23 |
|     1 Vendor Specific Element | 29681 | 8.72 |
|     2 Vendor Specific Element | 47375 | 13.92 |
|     3 Vendor Specific Element | 8661 | 2.54 |
|     4 Vendor Specific Element | 104559 | 30.72 |
|     5 Vendor Specific Element | 4525 | 1.33 |
| WPS – UUID-E | 35908 | 10.55 |
| Total Collected Probe Requests | 340360 | |

**Table. 1:** Probe request fields used to create device fingerprint and frequency of occurrence in data collected in our lab

*Device identification*

Combining the use of non-randomized MAC addresses, device fingerprint elements, use of transmitted SSIDs to differentiate devices, and UUID-E available in the probe requests with WPS field, we have enough information to identify a single Wi-Fi scan instance (Algorithm 1) as well as reappearance of a device. Even with MAC randomization, the information elements in the probe requests allow the adversary to identify devices.

After identifying the Wi-Fi scan instances, we start with device identification. Here we first check if the MAC addresses of 2 separate instances are the same. If they are we can consider the instances as the same device. If the MAC addresses are randomized or different from each other, we check for the presence of WPS fields, and subsequently check the UUIDE field and evaluate if the device is the same or not. In case the WPS field is not included and MAC addresses are not matching, we determine the similarity using the information elements section of probe requests and calculate a similarity score between the two preferred network lists. If the similarity is higher than a set threshold we can consider the scan instances to be from the same device. Since the preffered network list revealed through probe requests is in majority of the cases quite short and in many cases can be incomplete, the threshold was set to >0.5. Since with two transmitted SSIDs in each scan instance, one identical SSID will result in the similarity of 0.5. And since Wi-Fi networks have quite unique names, we take devices with at least two matching SSIDs as similar devices. The process of identifying a single device is shown in Algorithm 2.

*Temporal Analysis*

One of the more difficult parameters to mask for a single device sending multiple probe requests is the time difference between 2 probe requests. From our analysis of the Sapienza Probe Request dataset created by Barbera et. al (2013), slightly more than 98 % of subsequent probe requests sent by a single device are transmitted less than 65 milliseconds apart. These bursts of transmitted probe requests can be used for fingerprinting of the device. This is useful in conjunction with incrementing sequence numbers to distinguish two different devices and will be a potential threat to the users in the future, since the incrementing sequence number could reveal one device using multiple MAC addresses after every probe request.

We did not use the time difference between two probe requests, since devices do not change their MAC address during the scan instance. Instead we used different approach to time analysis. We used all of the similar device data that we got during device identification and analysed the recurrent appearances of each device and possible similarity to others. This way we discovered a pattern that allowed us to identify cases where single device looked like several devices. We did this by considering scan instance appearances of one device and clustering them together based on time. We then compared the number of clusters between devices. In case two devices had the same amount of appearance clusters, we checked the overlay between clusters. Subsequently we decided if the devices were in the end single device misidentified as many, or skip it it and move to the next device.

**Algorithm 1** Scan Instance Identification

```
 1:  variables
 2:      probe.mac, MAC address of the probe request
 3:      probe.has_wps, Probe request with WPS field
 4:      probe.uuide, UUID-E of the probe request
 5:      probe.ie, Information Element of the probe request
 6:      probe.sn, Sequence number of the probe request
 7:  end variables
 8:  if probe₁.mac = probe₂.mac then
 9:      if probe₁.has_wps and probe₂.has_wps then
10:          if probe₁.uuide = probe₂.uuide then
11:              return True                    ▷ True - Same instance
12:          else
13:              return False                   ▷ False - Different instance
14:          end if
15:      end if
16:      if probe₁.ie = probe₂.ie then
17:          if probe₁.sn < probe₂.sn < probe₁.sn + 5 then
18:              return True                    ▷ True - Same instance
19:          else
20:              return False                   ▷ False - Different instance
21:          end if
22:      else
23:          return False                       ▷ False - Different instance
24:      end if
25:  else
26:      return False                           ▷ False - Different instance
27:  end if
```

**Algorithm 2** Device Identification

```
 1:  variables
 2:      instance.mac, MAC address of the instance
 3:      instance.has_wps, Instance with WPS field
 4:      instance.uuide, UUID-E of the instance
 5:      instance.ie, Information Element of the instance
 6:      instance.SSIDs, List of SSIDs from one instance
 7:      threshold, Minimum similarity threshold
 8:  end variables
 9:  if instance₁.mac = instance₂.mac then
10:      return True                            ▷ True - Same device
11:  else if instance₁.has_wps and instance₂.has_wps then
12:      if instance₁.uuide = instance₂.uuide then
13:          return True                        ▷ True - Same device
14:      else
15:          return False                       ▷ False - Different device
16:      end if
17:  else if instance₁.ie = instance₂.ie then
```

18: $\quad p = \dfrac{set(instance_1.SSIDs) \text{ and } set(instance_2.SSIDs)}{set(instance_1.SSIDs) \text{ or } set(instance_2.SSIDs)}$

```
19:      if p > threshold then
20:          return True                        ▷ True - Same device
21:      else
22:          return False                       ▷ False - Different device
23:      end if
24:  else
25:      return False                           ▷ False - Different device
26:  end if
```

## Results

From the 340,360 probe requests collected at our office, we identified in total 125,983 scan instances. After following Algorithm 2, we got 1023 devices as is represented in Figure 2a). As a single instance we count any single probe request or burst of probe requests according to Algorithm 1. These instances were then clustered based on their similarity following the Algorithm 2. This way we were able to match at least two instances to a single device. If the tested instance showed no similarity to others, that instance was discarded as a single instance device that we had no way to track or to locate.

For devices that do not randomize their MAC addresses, the tracking is very effective and we can easily see when the device was inside of our office or in its proximity. Reason being, the unique identifier is the MAC address, which never changed. Due to this we were able to identify a significant number of devices that could be easily tracked and analysed for presence patterns. Between those devices were also a few that never left the proximity of the probe request sniffer as well as some that showed up in monitored range only for a few minutes. As presented in Figure 2a), 212 devices did not use MAC randomization. As an example comparison of presence in time for 8 such devices is shown in Figure 3.a). As a single instance we count any single probe request or burst of probe requests according to Algorithm 1. These instances were then clustered based on their similarity following the Algorithm 2. This way we were able to match at least two instances to a single device. If the tested instance showed no similarity to others, that instance was discarded as a single instance device that we had no way to track or to locate.

For devices that do not randomize their MAC addresses, the tracking is very effective and we can easily see when the device was inside of our office or in its proximity. Reason being, the unique identifier is the MAC address, which never changed. Due to this we were able to identify a significant number of devices that could be easily tracked and analysed for presence patterns. Between those devices were also a few that never left the proximity of the probe request sniffer as well as some that showed up in monitored range only for a few minutes. As presented in Figure 2a), 212 devices did not use MAC randomization and the example of presence in time for 8 devices is shown in Figure 3.

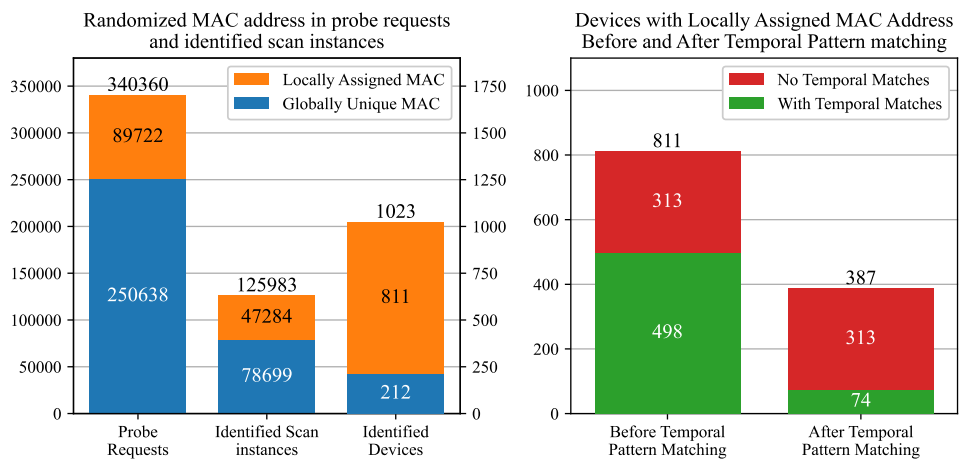

**Fig. 2:** Dataset information and device identification: a) Probe Requests, Identified Scan Instances and Identified Devices, b) Identified Devices before and after Temporal pattern Matching

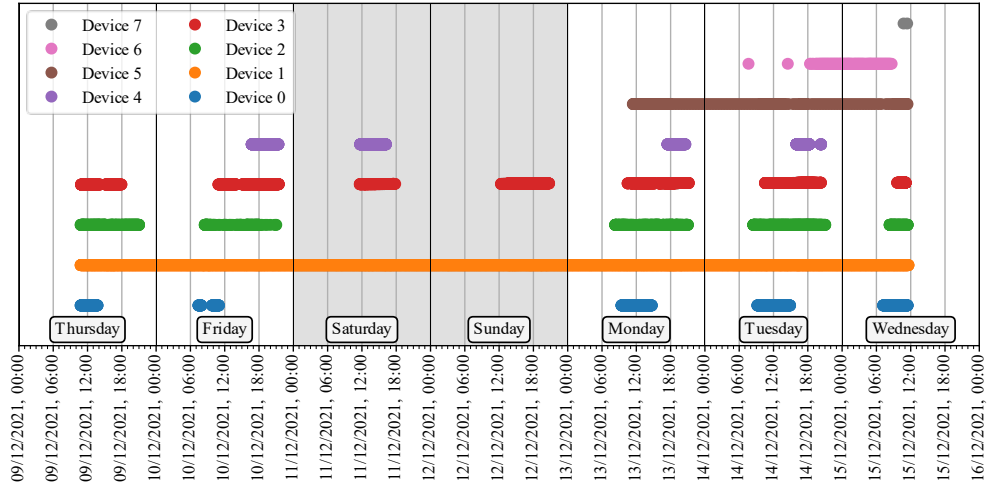

**Fig. 3:** Occurrence of several devices identified by the usage of globally unique MAC address

The identification of devices randomizing MAC addresses is more complicated, but despite MAC randomization making the process more difficult, we were able to identify many devices using the techniques mentioned before in the Analysis section. The results of our analysis can be seen on 8 devices using randomized MAC address in Figure 4. Even with the more complicated approach to identification, from the resulting data, the analysis of user presence is still possible.

The Algorithm 2 provides us with instances clustered as one device. The instance matching is not 100% accurate and in some cases, especially in those considering devices with randomized MAC addresses, can misidentify a single device as several devices. Using the Algorithm 2, we matched the 125,983 instances to 1023 devices, which can be seen in Figure 2a). After the instance matching, we used the temporal analysis we proposed on the identified devices. We managed to detect similarity in between 498 misidentified devices, and reduce this amount to only 74 devices using MAC randomization. 313 devices with locally assigned MAC addresses did not match temporal pattern of other devices. The amount of devices with locally assigned MAC address before and after temporal pattern matching can be seen in Figure 2b). The probe request transmission patterns were quite closely matching each other, as can be seen on Figure 5, which led us to identifying these appearances as a single device or single user carrying multiple devices.

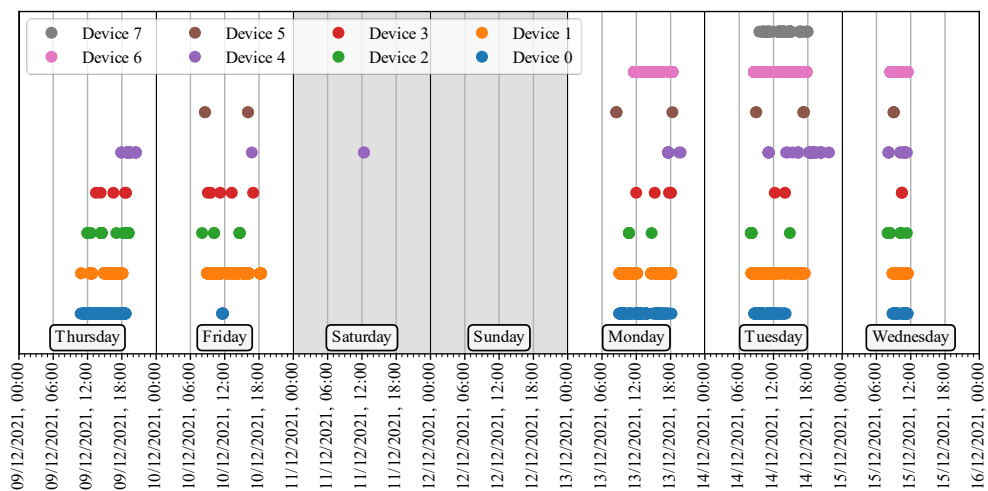

**Fig. 4:** Occurrence of several devices identified despite the use of MAC randomization

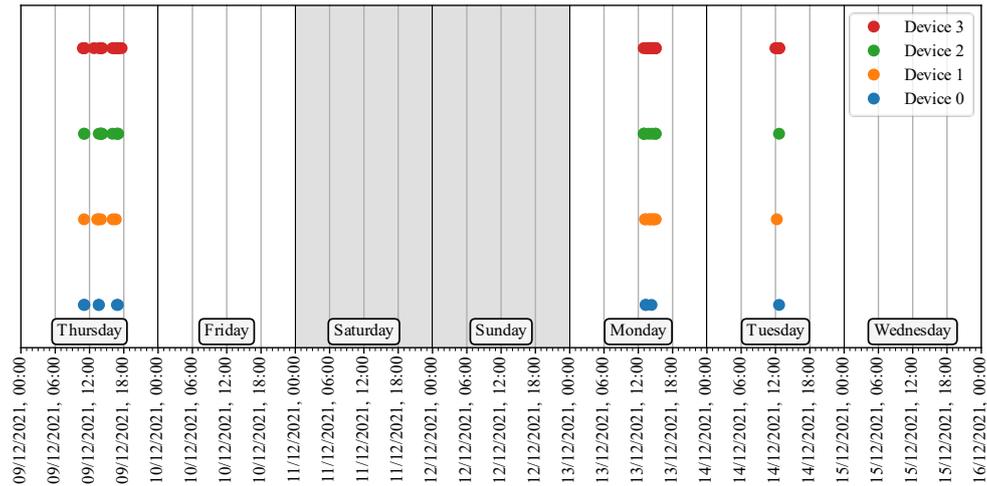

**Fig. 5:** Occurrence of single device misidentified as multiple devices, later identified as single device through the similarity in temporal patterns

**Conclusions and Future Work**

In this paper we explored the current state of privacy regarding probe requests in the 802.11 standard using our captured dataset of probe requests. Through deterministic methods we explored the possibilities to bypass and identify devices without the need for using the globally unique MAC address. From our results we managed to track many devices with and without locally assigned MAC addresses.

We also introduced an approach to use temporal pattern matching to identify device appearing as several devices due to the use of MAC address randomization. Using this trchnique we managed to reduce 498 identified devices to just 74. This makes the temporal pattern matching quite an effective technique for detecting devices despite using MAC address randomization.

For the future we plan to continue the exploration of privacy with probe requests and we plan to collect and publish a new probe request dataset. We also plan to release a small dataset without the use of anonymization techniques, from our controlled environment and with consent of everyone involved in the data capture.

**Acknowledgement**

The authors gratefully acknowledge funding from European Union's Horizon 2020 Research and Innovation programme under the Marie Sklodowska Curie grant agreement No. 813278 (A-WEAR: A network for dynamic wearable applications with privacy constraints, http://www.a-wear.eu/) and No. 101023072 (ORIENTATE: Low-cost Reliable Indoor Positioning in Smart Factories, http://orientate.dsi.uminho.pt). This work does not represent the opinion of the European Union, and the European Union is not responsible for any use that might be made of its content.